\documentclass[dvips]{jfm}

        \newif\ifdraft
\draftfalse     

\usepackage[dvips]{color,graphicx}
\usepackage{amsmath,amsfonts,amssymb}
\usepackage{natbib}
\usepackage{subfigure}

\newcommand {\norm} [1] { \lVert #1 \rVert}

    \ifdraft    

\newcommand{\PC}[1]{$\footnotemark\footnotetext{PC: #1}$}

\newcommand{\DV}[1]{$\footnotemark\footnotetext{DV: #1}$}

\newcommand{\JFG}[1]{$\footnotemark\footnotetext{JG: #1}$}

\newcommand{\JH}[1]{$\footnotemark\footnotetext{JH: #1}$}

    \else   

\newcommand{\PC}[1]{}
\newcommand{\JFG}[1]{}
\newcommand{\DV}[1]{}
\newcommand{\JH}[1]{}

    \fi

\newcommand{\ie}{{i.e.}}        

\newcommand{\statesp}{state space}

\newcommand{\eqva}{equilibria}

 \newcommand{\reqva}{traveling waves}

\newcommand{\po}{periodic orbit}

\newcommand{\bu}{\ensuremath{{\bf u}}}
\newcommand{\jEigvec}[1]{\ensuremath{{\bf e}^{(#1)}}}   

\newcommand{\uEQout}{\ensuremath{{\bf u}_{\text{\tiny out}}}}   
\newcommand{\uEQin}{\ensuremath{{\bf u}_{\text{\tiny in}}}}     

\newcommand{\EQV}[1]{\ensuremath{\text{EQ}_{#1}}}
\newcommand{\tLM}{\ensuremath{{\text{EQ}_0}}}
\newcommand{\tLB}{\ensuremath{{\text{EQ}_1}}}
\newcommand{\tUB}{\ensuremath{{\text{EQ}_2}}}
\newcommand{\tNNB}{\ensuremath{{\text{EQ}_3}}}
\newcommand{\tNB}{\ensuremath{{\text{EQ}_4}}}
\newcommand{\tEQfive}{\ensuremath{{\text{EQ}_5}}}

\title[Heteroclinic Connections in Plane Couette Flow]
{Heteroclinic connections in plane Couette flow}

\author[
J. Halcrow,
J. F. Gibson,
P. Cvitanovi\'c
    and
D. Viswanath
]{
$^1$J.\ns  H\ls A\ls L\ls C\ls R\ls O\ls W
,\ns
$^1$J.\ns F.\ns G\ls I\ls B\ls S\ls O\ls N
,\ns
$^1$P.\ns C\ls V\ls I\ls T\ls A\ls N\ls O\ls V\ls I\ls \'C
\and
$^2$D.\ns  V\ls I\ls S\ls W\ls A\ls N\ls A\ls T\ls H
}

\affiliation{
$^1$School of Physics,
Georgia Institute of Technology,
Atlanta, GA  30332, USA\\
[\affilskip]
$^2$Department of Mathematics,
University of Michigan,
Ann Arbor, MI 48109, USA}

\begin{document}
\maketitle

\begin{abstract}
Plane Couette flow transitions to turbulence for $Re\approx 325$
even though the laminar solution with a linear profile is linearly
stable for all $Re$ (Reynolds number). One starting point for
understanding this subcritical transition is the existence of
invariant sets in the {\statesp} of the Navier Stokes equation,
such as upper and lower branch equilibria and periodic and relative
periodic solutions, that are quite distinct from the laminar
solution.
This article reports several heteroclinic connections between such
objects and briefly describes a numerical method for locating
heteroclinic connections. Computing such connections is
essential for understanding the global dynamics of spatially localized
structures that occur in transitional plane Couette flow. We show that
the nature of streaks and streamwise rolls can change significantly
along a heteroclinic connection.
\end{abstract}

\vspace*{-0.6cm}
\section{Introduction}
In plane Couette flow, the fluid between two parallel walls of fixed
separation
is driven by the motion of the walls in opposite directions.
Even though the laminar solution is linearly stable for
all $Re$ (Reynolds number) as shown by \cite{KLH},
turbulent spots
evolve into large turbulent patches for $Re$ exceeding the modest
value of about $325$ \citep{BDMD}. These turbulent patches are
sustained by the flow for very long, and possibly infinite, time
intervals. 
From a dynamical point of view, the evolution of the
velocity field corresponds to a trajectory in {\statesp}, and
indefinitely sustained motion should correspond to {\it invariant}
sets.  Invariant sets in {\statesp} have the property that a
trajectory that starts exactly on such a set stays on that set
forever, and a trajectory that starts outside that set cannot
land on it within a finite time interval although it can approach the
invariant set rapidly. Thus a reasonable starting point for understanding
when and why turbulence becomes sustained in plane Couette flow and
other shear flows is not the loss of linear stability of laminar flow,
which never happens in plane Couette flow, but the existence of
invariant sets. Equilibria, traveling waves, periodic solutions,
and relative periodic solutions are all invariant sets. The union
of such sets can form chaotic saddles or chaotic attractors,
invariant sets which may explain a good deal of the
dynamics of shear flows \citep{SE}. Thus the numerical computation
of equilibria, traveling waves, periodic solutions, and relative periodic
solutions
\citep{Nagata1,Nagata2,W03,Viswanath1,GHC,HGC08,GHC08a}
is a step
towards understanding the dynamics of plane Couette flow in the
transitional regime.

We use a computational box of extent $0\leq x \leq 2 \pi/\alpha$,
$-1\leq y \leq 1$, and $0\leq z \leq 2\pi/\gamma$, with $\alpha =
1.14$ and $\gamma=2.5$ \citep{W03}, where $x$, $y$, $z$ are the
streamwise, wall-normal, and spanwise coordinates,
respectively. Likewise, $u$, $v$, $w$ are the three components of the
velocity field.  The boundary condition is periodic along $x$ and $z$,
and no-slip at the walls.
For comparison, the experimental setup of \cite{BDMD}
is about a meter long with a separation between the walls of only $7$
mm. At the moment, small computational boxes are needed to keep the
cost of computing invariant sets manageable.  Nevertheless, small
computational boxes are capable of picking up significant aspects of
turbulent boundary layers and transitional dynamics, perhaps because
some of the features of those regimes are localized in space.  For
instance, periodic and relative periodic orbits computed in a small
box reproduce the formation and break-up of streaks in the near-wall
region \citep{Viswanath1}. Indeed, such solutions show that the
spanwise advection of coherent structures could be a significant
source of the spanwise variation of the root mean square value of the
streamwise velocity. In addition, turbulent spots, which appear to
play a key role in transition, are localized in space, although
certain aspects of their dynamics depend upon the interaction with the
external flow \citep{SchuE}.

In this article, we report three heteroclinic connections between
equilibrium (or steady) solutions of plane Couette flow at $Re=400$,
where the $Re$ is based on half the difference in velocity between the
moving walls, half the distance between the walls, and the kinematic
viscosity of the fluid.  Basic data for six such solutions, with the first one
being the laminar solution, is given by Table~\ref{table-1}
(for detailed data sets the reader can consult
{\tt Channelflow.org} and \cite{HalcrowThesis}). The
equations of plane Couette flow are unchanged by the shift-reflect and
shift-rotate transformations defined in Section 3. All the equilibria
lie in the $S$-invariant subspace, which is the space of velocity
fields invariant under both transformations.  The equilibria $\tLB$
and $\tUB$ are called lower and upper branch solutions
\citep{Nagata1,W03}.
The heteroclinic connections reported here are from
$\tNNB$, $\tNB$, and $\tEQfive$ to $\tLB$; and from
$\tLB$ to $\tUB$.

\begin{table}
\centering
\begin{tabular}{c|ccccccc}
 & $I=D$ & $E$ & $E_{roll} / E$ & $d(W^u)$ & $d(W^u_S)$ & $\lambda_0$
& $Re_\tau$\\
\hline
$\tLM$ & 1 & 1 & 0 & 0 & 0                           & -0.00616850
& $40$\\
$\tLB$   & 1.429258  &  0.817778  &  0.000330 & 1 & 1 & 0.05012078
& $47.82$\\
$\tUB$   & 3.043675  &  0.468225  &  0.018323 & 8 & 2 & 0.05558362
& $69.78$\\
$\tNNB$  & 1.317683  &  0.829378  &  0.000759 & 4 & 2 &
$0.03397837 \pm 0.01796294\,\imath$ & $45.92$\\ 
$\tNB$ & 1.453682  &  0.746056  &  0.002515 & 6 & 3 & 0.03064964
& $48.23$\\
$\tEQfive$ & 2.020135&  0.644223 &  0.003511 & 11 & 4 &
$0.07212161 \pm 0.04074989\,\imath$ & $56.85$\\
\end{tabular}
\caption{
 Basic statistics for equilibria at  $Re=400$.
 $\tLM$ is the laminar solution of plane Couette flow.
  The
 rate of energy input $I$ and  the rate of dissipation $D$
 are both  normalized
 to be $1$ for the laminar state.  So is the
 total kinetic energy denoted by $E$.
 $E_{roll} / E$ is the fraction of the total kinetic energy in the rolls.
 The dimension of the unstable manifold is
 $d(W^u)$, while $d(W^u_S)$ is the
 dimension of the intersection of the unstable manifold with the
 $S$-invariant subspace. $\lambda_0$ is the eigenvalue with the
 greatest real part, and
 $Re_\tau$ is the width of the channel in wall units.}
\label{table-1}
\end{table}

In the presence of continuous rotation symmetry and discrete reflection
symmetry, the existence of heteroclinic cycles follows
from the normal form of certain codimension-2 bifurcations \citep{Kuznetsov}.
\cite{ALMP1, ALMP2} have shown that Taylor-Couette flow with
a stationary outer cylinder undergoes a codimension-2 bifurcation, the
normal form of which implies the existence of a heteroclinic cycle.
That the basic laminar solution of this Taylor-Couette flow undergoes
a sequence of supercritical bifurcations, making it possible to track
bifurcations while computing only linearly stable solutions, while the
transition in plane Couette flow is subcritical is just one difference
from our work.  Notably, the computations of \cite{ALMP1} use a domain
and boundary conditions that match their experimental setup.  We do
not compute codimension-2 bifurcations, although we return to that
point and the influential thesis of \cite{Schmiegel} in Section
4. In addition, our computations of heteroclinic connections
are explicit and make use of the
eigenvalues and eigenvectors of the linearizations around the
equilibria.

Instead, our computations rely on the simple principle that an object
of dimension $k$ is likely to intersect
in a stable way an object whose codimension in
{\statesp} is less than or equal to $k$. At the
bottom, this is nothing more than the fact that two
submanifolds in
general position can intersect if the sum of their dimensions is greater
than or equal to the dimension of the \statesp
(whether they actually intersect is a subtle question that is
central to the
``structural stability'' of ergodic dynamical systems
\citep{smale}).
For an illustration in the nonlinear setting, see \cite{ASBook}.
\cite{KNS} (see Section 5 of their paper)
make elegant use of this principle and of
invariant subspaces implied by discrete symmetries
of the underlying
PDE to numerically deduce the existence of a heteroclinic
connection
in the Kuramoto-Sivashinsky equation. Indeed, they comment that their
work may have implications for shear flows. With regard to the
heteroclinic connections presented here, it is significant to note
from Table~\ref{table-1} that the codimension of the stable manifold
in the $S$-invariant space (which is equal to $d(W^u_S)$) of $\tLB$
is less than the value of $d(W^u_S)$ for $\EQV{i}$ with $i=3,4,5$.
Thus it is not surprising that the unstable manifolds of $\EQV{i}$
with $i=3,4,5$ intersect the stable manifold of $\tUB$
in a stable way (\ie, robustly with respect to small changes of
system parameters).

All the equilibria in Table~\ref{table-1}, except $\tEQfive$, have
well-formed streaks, which means that the streamwise velocity has
pronounced variation in the spanwise direction. The streaks are
accompanied by streamwise rolls which is the typical situation for
boundary layers \citep{KKR}. Streaks and streamwise rolls are also
found near the edges of turbulent spots \citep{DD, Tillmark, SchuE}.
They could be relevant to the wavelike manner in which the turbulent
spots spread to form patches.
In this regard, we note that heteroclinic
connections are important to obtaining a global picture of the dynamics in
{\statesp}. They can be useful for the physical space picture as
well, as shown by the dramatic change in the balance between rolls and
streaks along the heteroclinic connection from
$\tEQfive$ to $\tLB$. In Section 3, we
present a {\statesp} plot in the manner of \cite{GHC} to show how the
heteroclinic connections at $Re=400$ are related to one another.

\vspace*{-0.6cm}
\section{Finding and verifying heteroclinic connections}
The discretization of the computational box used $32$ Fourier
points in the $x$ direction, $35$ Chebyshev points in the $y$ direction,
and $32$ Fourier points in the $z$ direction. Direct numerical simulation
of plane Couette flow was performed using {\tt Channelflow.org} \citep{Gibson}.
The equilibria listed in Table~\ref{table-1} were found using
GMRES-hookstep iterations \citep{Viswanath1}. If the velocity fields
of the equilibria are integrated for a certain fixed time, they are nearly
unchanged. Yet the evolution of perturbations under such an integration
can be used along with the Arnoldi iteration to determine
all unstable eigenvalues and
eigenvectors, as well a set of the least contracting stable
eigenvalues and eigenvectors
\citep{Viswanath1}. Such a computation was used to produce
the information about the unstable manifolds of the equilibria listed
in Table~\ref{table-1}.

The shift-reflect and shift-rotate transformations of a velocity
field are given by
\begin{align}
s_1 [u,v,w](x, y, z) &= [u, v, -w](x + L_x/2, y, -z), \nonumber\\
s_2 [u,v,w](x, y, z) &= [-u, -v, w](-x+L_x/2, -y , z +L_z/2),
\label{eqn-2-1}
\end{align}
respectively, where $L_x$ and $L_z$ are the periods of the
computational box in the $x$ and $z$ directions. If either
transformation is applied to a trajectory of plane Couette flow, one
gets another trajectory of plane Couette flow. The space of velocity
fields unchanged by both $s_1$ and $s_2$ is an invariant subspace
called the $S$-invariant space in \cite{GHC}. All the computations in this
paper are restricted to this invariant space. The norm used over
velocity fields of plane Couette flow throughout this paper is the
square root of the kinetic energy, with a normalization that makes the
norm of the laminar solution equal to $1$.

In a heteroclinic connection, the velocity field of plane Couette flow
varies over a time (or $t$) interval infinite in both senses, approaching
equilibria as $t\rightarrow -\infty$ and as $t\rightarrow \infty$.
Those are the initial and final equilibria of the heteroclinic
connection.
Since it is impossible to integrate over an infinite time interval,
our computed heteroclinic connections
start out in the linearized neighborhood and close to
the initial or ``out'' equilibrium \uEQout, and end
in the linearized neighborhood close to the final or
``in'' equilibrium \uEQin, after a
finite interval of time.
For the heteroclinic
connections
that go from
$\tNNB$ and $\tEQfive$ to $\tLB$, the initial point on the computed
heteroclinic connection is a perturbation using the two dimensional
eigenspace that corresponds to the complex pair of eigenvalues with
the greatest real part.
It is reasonable to look in that space because all except a proper
subspace of trajectories that originate near an equilibrium point
are tangent to the leading eigenspace, which is $1$-dimensional
if the eigenvalue with the largest real part is real and simple
and $2$-dimensional if a simple complex pair. For the same reason,
it is perhaps surprising that the heteroclinic connection from
$\tNB$ to $\tLB$ is found using an eigenspace corresponding to
a complex pair with real part less than that of one other eigenvalue.

\begin{figure}
\begin{center}
(a)
\includegraphics[height=1.5in,width=1.5in]{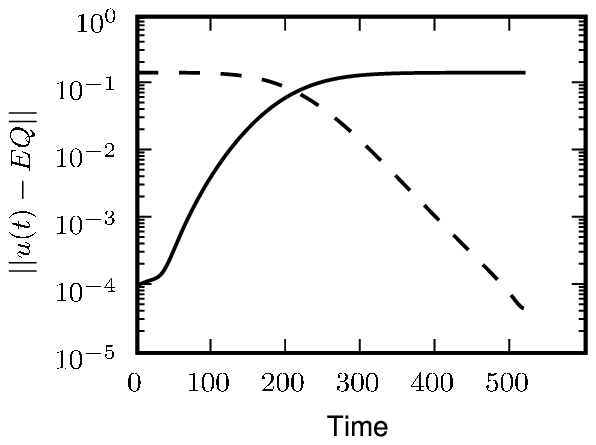}
(b)
\includegraphics[height=1.5in,width=1.5in]{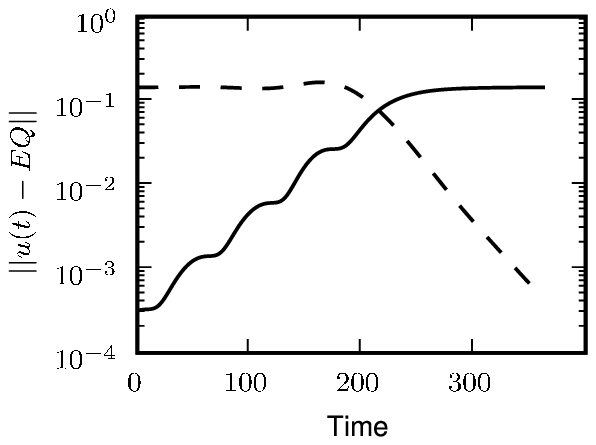}
(c)
\includegraphics[height=1.5in,width=1.5in]{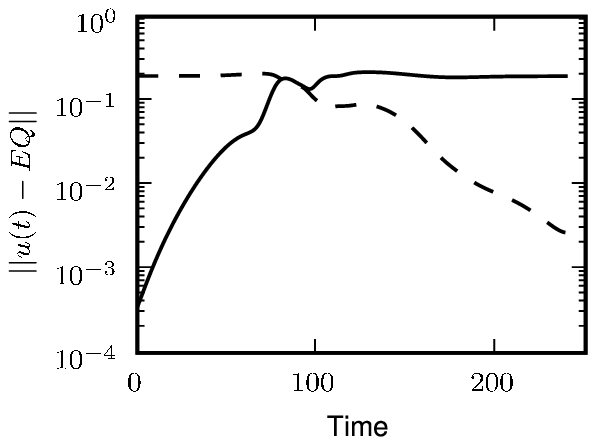}
\end{center}
\caption[xyz]{
Plots of distances from the initial (solid line) and final (dashed
line) equilibria to the velocity field at varying times along the
computed heteroclinic connection. (a), (b), (c) correspond to the
heteroclinic connections into $\tLB$ from $\tNNB$, $\tNB$, and $\tEQfive$,
respectively.}
\label{fig-1}
\end{figure}

Let $\jEigvec{1}$, $\jEigvec{2}$ be an orthonormal basis for the
two-dimensional eigenspace that corresponds to a complex eigenvalue
pair of the equilibrium $\uEQout$.  The span will be tangent to the
unstable manifold at the equilibrium.  We consider the set of velocity
fields of plane Couette flow that at the initial time $T=0$ lie on 
a circle of radius $r$:
\begin{equation*}
\bu(0)_\phi = \uEQout + r (\jEigvec{1} \cos \phi + \jEigvec{2} \sin \phi)
\,.
\end{equation*}
For a small and fixed value of $r$,
we search for a point on this circle which evolves to make the closest approach
to another equilibrium, \uEQin.
Let 
\begin{equation*}
 G(\phi) = \min_{T} \norm{\bu(T)_\phi - \uEQin},
\end{equation*}
where $\bu(T)_\phi$ is the velocity field that results from
evolving the velocity field $\bu(0)_\phi$ for time $T$ and where the
minimizing value of $T$ is the time of the first local minimum greater
than a certain threshold. 
The closest approach is the minimum of $G(\phi)$ over $0\leq \phi < 2\pi$.
Since $G(\phi)$ is a function of a single
real variable, it can be minimized using any one of a number of
well-known and effective methods. The computation of heteroclinic
connections sketched above uses a first order asymptotic boundary
condition at the initial equilibrium. For small systems, it is possible to
use an asymptotic boundary condition at the final equilibrium as well.
For an example, see \cite{DDF}.

For the computed heteroclinic connections from $\tNNB$, $\tNB$, and
$\tEQfive$ to $\tUB$, the chosen values of $r$ were 0.0001, 0.0003, and 0.0004,
respectively.  Figure~\ref{fig-1} shows data for the three computed
heteroclinic connections. In each plot of that figure, the solid line
is tiny at the beginning but rises exponentially while the dashed line
is flat. Therefore, we conclude that the initial part of each computed
heteroclinic connection is in a region where its time evolution is
governed by the linearization around its initial equilibrium. Similarly,
we can conclude that the final part is in a region where the evolution is
governed by the linearization around the final equilibrium.

To verify the computed heteroclinic connections using another code, it
could be necessary to use three stages. The computed connection from
$\tEQfive$ to $\tLB$, for instance, spends about $75$ time units near
the initial equilibrium and more than $100$ units near the final
equilibrium, as evident from Figure~\ref{fig-1}.  Using the data in
Table~\ref{table-1}, one may easily estimate that the loss of
precision in those two stages is more than $3$ digits. As the
equilibria themselves are computed with only about $4$ or $5$ digits
of precision, one has to do the verification in segments.  Such a
verification of Figure~\ref{fig-1}, which was performed using a
completely independent code \citep{Viswanath1}, and the applicability
of shadowing theorems about numerical trajectories
\citep{Palmer}
leave little doubt that the computed
heteroclinic connections are real.

\begin{figure}
\begin{center}
(a)
\includegraphics[height=1.5in,width=1.5in]{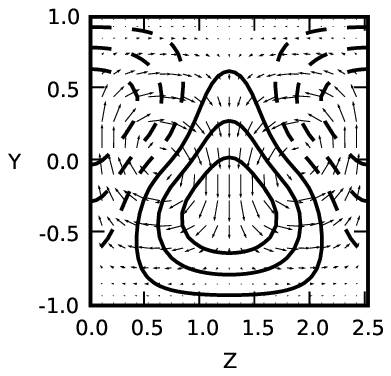}
(b)
\includegraphics[height=1.5in,width=1.5in]{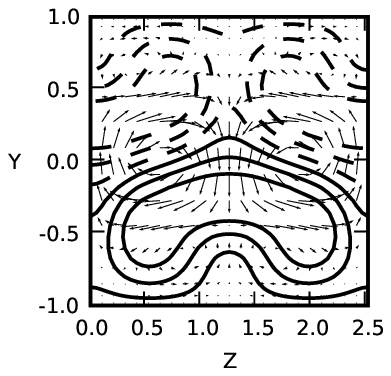}
(c)
\includegraphics[height=1.5in,width=1.5in]{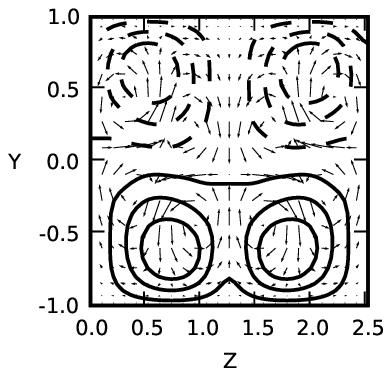}\\
(d)
\includegraphics[height=1.5in,width=1.5in]{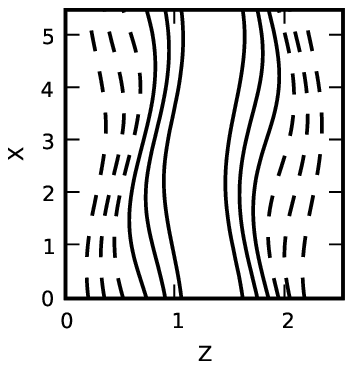}
(e)
\includegraphics[height=1.5in,width=1.5in]{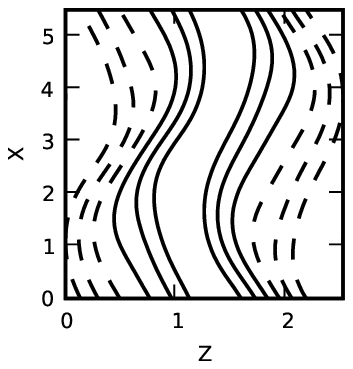}
(f)
\includegraphics[height=1.5in,width=1.5in]{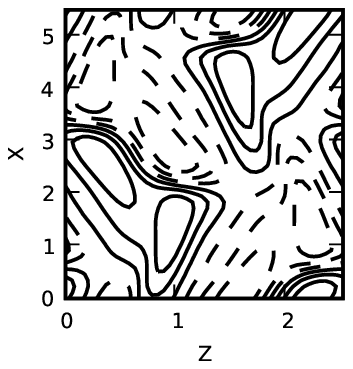}
\end{center}
\caption[xyz]{(a), (b), (c)
correspond to $\tLB$, $\tNB$, $\tEQfive$, while
(d), (e), (f) correspond
to $\tLB$, $\tNNB$, $\tEQfive$, respectively (there is very little
difference in the plots for $\tNNB$ and $\tNB$).  The quiver plots in the top
row show the streamwise averaged velocity components $v$ and $w$ in
the $y-z$ plane. The six  contour lines of the streamwise averaged
$u$ component
are equispaced in $(u_{max},-u_{max})$, with $u_{max}$ being $0.44$,
$0.27$ and $0.45$ for $\tLB$, $\tNB$ and $\tEQfive$, respectively,
and with the negative lines being dashed. The bottom plots show
six contour lines of $u$ in the section $y=0$. The contour lines
are equispaced in $(u_{max}, -u_{max})$, with $u_{max}$ being
$0.33$, $0.22$ and $0.18$ for $\tLB$, $\tNNB$ and $\tEQfive$, respectively,
and with the negative lines being dashed.
lots would be better, maybe a stride of 2 in both $x$ and $y$.]
}
\label{fig-2}
\end{figure}

\vspace*{-0.6cm}
\section{Heteroclinic connections at $Re=400$}
The top plots of Figure~\ref{fig-2} show the correlation between the
rolls and the position of the streaks.
The equilibria
$\tLB$, $\tNNB$ and $\tNB$ each have a single counter-rotating pair of rolls.
The rolls
distort the mean flow, which increases with $y$, and thus
explain the position
of the streaks in Figures \ref{fig-2}a and b \citep{Kerswell}.
$\tEQfive$ has four counter-rotating pairs. From Figure~\ref{fig-2}f,
we see that the mid-plane flow is not at all streaky for $\tEQfive$.

\begin{figure}
\begin{center}
(a)
\includegraphics[height=1.5in,width=1.5in]{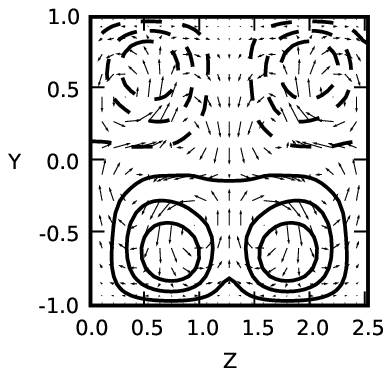}
(b)
\includegraphics[height=1.5in,width=1.5in]{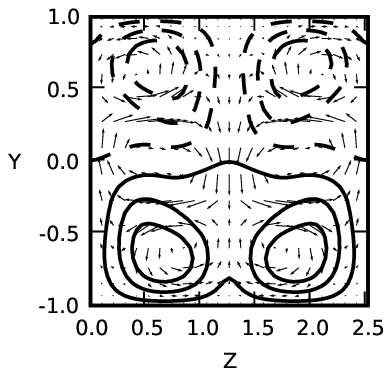}
(c)
\includegraphics[height=1.5in,width=1.5in]{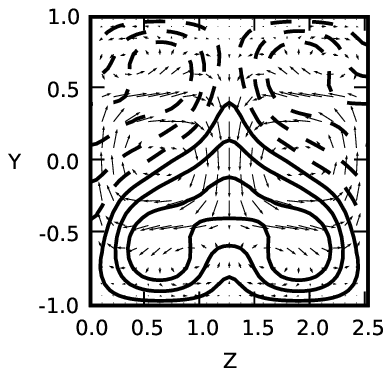}\\
(d)
\includegraphics[height=1.5in,width=1.5in]{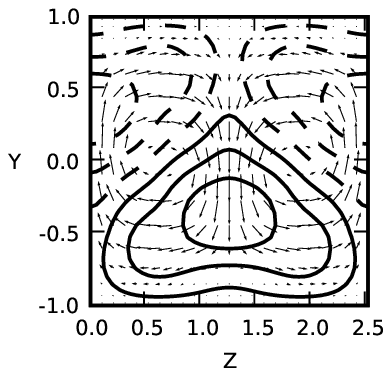}
(e)
\includegraphics[height=1.5in,width=1.5in]{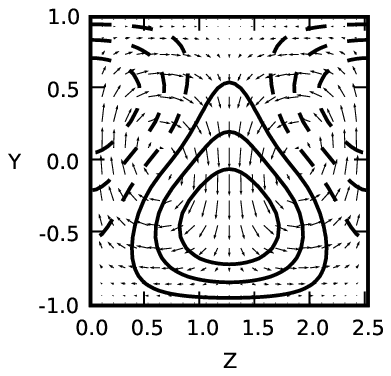}
\end{center}
\caption[xyz]{
(a), (b), (c), (d), (e) are plots of the velocity field at $t=50$, $t=75$,
$t=90$, $t=100$, and $t=150$, respectively, of the computed heteroclinic connection
from $\tEQfive$ to $\tLB$
of Figure~\ref{fig-1}.
The plots are similar to the ones in the top
row of Figure~\ref{fig-2}, with values of $u_{max}$ being $0.48$, $0.44$, $0.36$,
$0.46$ and $0.53$, respectively.
}
\label{fig-3}
\end{figure}

Figure~\ref{fig-3} illustrates the manner in which the rolls change in
form along the heteroclinic connection from $\tEQfive$ to $\tLB$. From
Figure~\ref{fig-1}c, it is evident that for $t\in [75,125]$ the
computed heteroclinic connection does not follow the linearized
dynamics around its initial or final equilibrium. Figure~\ref{fig-3}
confirms that the rolls change in form within that interval. While
the coexistence of rolls and streaks in turbulent boundary layers
is well known \citep{KKR}, the sort of coalescence of rolls that
is observed in Figure~\ref{fig-3} is a new type of behavior.

\begin{figure}
\begin{center}
\includegraphics[width=0.8\textwidth]{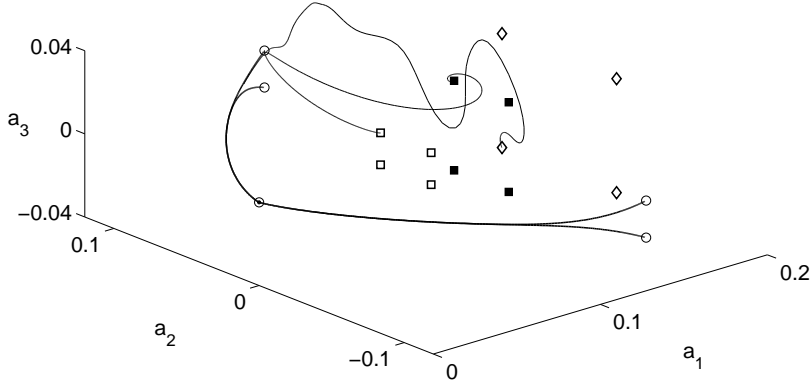}
\end{center}
\caption{In the state space plot of the three heteroclinic connections
at $Re=400$ above, $\tNNB$, $\tNB$, and $\tEQfive$, along with their images
under the half shifts $\tau_x$, $\tau_z$ and $\tau_{xz}$, are denoted by
the symbols {\mbox{\scriptsize $\square$}},\mbox{\scriptsize $\blacksquare$}, and
$\lozenge$
respectively. The laminar solution
$\tLM$ ($\odot$)
is at the origin. $\tUB$ and its images under the half shifts are denoted
by $\mbox{\Large $\circ$}$. The meaning of the axes is explained in the text.
            }
\label{fig-4}
\end{figure}

The significance of the heteroclinic connections is that
they give a  global picture of the dynamics, a  picture that cannot
can be inferred from equilibria alone. To visualize global
dynamics, it is essential to depict the equilibria and the heteroclinic
connections between them in {\statesp}. The {\statesp} of plane Couette
flow is infinite dimensional, and in the spatial discretization used
for computing the heteroclinic connections, it is more than $6\times 10^4$,
which is still much too large. Figure~\ref{fig-4} uses
a 3-dimensional projection
of points in that {\statesp}, which was introduced
by \cite{GHC}, to depict the equilibria and
the known connections between them.

Before explaining the projections used in Figure~\ref{fig-4}, we
discuss why that figure can be considered a good visualization of the
known heteroclinic connections of plane Couette flow at $Re=400$.  It
is typical to use projections to construct low dimensional models and
these models are considered reasonable if they capture $90\%$ of
the energy in the underlying flow, for instance. Such
models use many more dimensions than just three axes, which is all
that can be used in a depiction such as Figure~\ref{fig-4}. In addition,
if the projected velocity field has $90\%$ of the energy, its
normwise relative error can be as high as $30\%$. For these reasons,
we do not use the amount of energy retained by the projection to judge
the quality of depictions such as Figure~\ref{fig-4}.
       
Instead, we adopt a more geometric point of view. If a curve in a
Hilbert space is projected onto a finite dimensional plane, it
develops artificial cusps or corners at points where the
tangent to the curve
is orthogonal to
the plane of projection. For instance,
the projection of a smooth curve in
$R^3$ to the normal plane at a fixed point on the curve has an
artificial cusp \citep[chapter 3]{Widder}. In Figure~\ref{fig-4}, we
see that the heteroclinic connections can be wavy but do not have
cusps. It is significant that the same projection gives a good
depiction of all the three heteroclinic connections into $\tLB$ and
the heteroclinic connection from $\tLB$ to the laminar solution, which
is shown as a thick line.

To explain the coordinates $a_1, a_2, a_3$ in Figure~\ref{fig-4},
we define $\tau_x$ and $\tau_z$ as follows:
\begin{align*}
\tau_x [u,v,w](x, y, z) = [u, v, w](x + L_x/2, y, z), \\
\tau_z [u,v,w](x, y, z) = [u, v, w](x, y , z +L_z/2),
\end{align*}
where $L_x$ and $L_z$ are the periods of the computational box
along $x$ and $z$. In addition, $\tau_{xz} = \tau_x \tau_z$.
For each equilibrium that is in the $S$-invariant subspace
\eqref{eqn-2-1}, one may apply
$\tau_x$, $\tau_z$, and $\tau_{xz}$ to get three other equilibria
that lie in the $S$-invariant subspace. In the same manner,
one may use each computed heteroclinic connection to get three others.
Only a single copy of each is shown in Figure~\ref{fig-4}.

Let $\hat{\bf u}_2$ be the velocity field of the upper-branch solution
$\tUB$, with the laminar velocity field subtracted. If
$\jEigvec{i}$ are defined by
\begin{align*}
\jEigvec{1} &= c_1 (1 + \tau_x + \tau_z + \tau_{xz})\hat{\bf u}_2\\
\jEigvec{2} &= c_2 (1 + \tau_x - \tau_z - \tau_{xz})\hat{\bf u}_2\\
\jEigvec{3} &= c_3 (1 - \tau_x + \tau_z - \tau_{xz})\hat{\bf u}_2\\
\jEigvec{4} &= c_4 (1 - \tau_x - \tau_z + \tau_{xz})\hat{\bf u}_2,
\end{align*}
with $c_i$ being normalizing constants, the $\jEigvec{i}$ form
an orthonormal set \citep{GHC}. For a given velocity field of
plane Couette flow, the $a_i$ are obtained by subtracting the
laminar flow from the velocity field and then taking the inner
product with $\jEigvec{i}$, where $i=1,2,3,4$.

The use of the upper branch equilibrium $\tUB$ to define $\jEigvec{i}$
and $a_i$ 
may appear arbitrary and to an extent
it is. Heuristically it is a good choice because the computations of
\cite{GHC} show that the dynamics of plane Couette flow, including
turbulent episodes and trajectories that relaminarize quickly, appear
to be trapped between the unstable manifolds of $\tUB$ and its three
images obtained by applying $\tau_x$, $\tau_z$ and $\tau_{xz}$
and the laminar solution.

\begin{table}
\begin{center}
\begin{tabular}{c|cccccc}
EQ & $I=D$ & $E$ & $E_{roll} / E$ & $d(W^u)$ & $d(W^u_S)$ & $\lambda_0$ \\
\hline
$\tLM$   & 1 & 1 & 0 & 0 & 0                          & -0.010966 \\
$\tLB$   & 1.710086  &  0.722516  &  0.002526 & 3 & 1 & 0.02524949 \\
$\tUB$   & 2.076045  &  0.634025  &  0.006357 & 4 & 2 & 0.0441718 \\
\end{tabular}
\end{center}
\caption[xyz]{The columns have the same meaning as in Table~\ref{table-1},
but with the equilibria computed at $Re=225$.}
\label{table-2}
\end{table}

\vspace*{-0.6cm}
\section{A heteroclinic connection at $Re=225$}
Table~\ref{table-2} gives data for $\tLM$, $\tLB$ and $\tUB$ at $Re=225$.
By comparing the dimensions of the unstable manifolds and their
restrictions to the $S$-invariant space, we can infer that both
$\tLB$ and $\tUB$ undergo bifurcations as $Re$ is increased from
$225$ to $400$. The dimension of $\tLB$'s unstable manifold is just
$1$.
By following that unstable manifold,
we found a heteroclinic connection to $\tUB$.

While the dimension of $\tLB$'s unstable manifold in the $S$-invariant
subspace is $1$, the codimension of $\tUB$'s stable manifold in the
same subspace is $2$. Based on that consideration alone a heteroclinic
connection seems implausible. However, this heteroclinic connection
is very likely related to a codimension-2 bifurcation. In such a
scenario, the dimensions of the unstable manifold of the initial
equilibrium and of the stable manifold of the final equilibrium must
be compared only within the center manifold.

\cite{Schmiegel} has systematically studied bifurcations of the solutions
of plane Couette flow found by \cite{Nagata1} and \cite{CB} using
a representation with about $1200$ modes. He has found heteroclinic
connections where the saddle node bifurcation that gives rise to
$\tLB$ and $\tUB$ is followed soon after by a pitchfork bifurcation
as $Re$ is increased. The heteroclinic connection reported above
is probably of that type.

To understand this heteroclinic connection better, it could be useful to
think of $L_z$, the spanwise size of the computational box, as
a parameter. In the parameter space with $Re$ and $L_z$ as the axes,
the saddle-node bifurcations that give rise to $\tLB$ and $\tUB$
will form a curve. There will be another curve that corresponds to
the pitchfork or the Hopf bifurcation. At the intersection of those
curves, we will have a codimension-2 bifurcation. An advantage of
realizing a heteroclinic connection using the normal form of a
codimension-2 bifurcation is that we will get a heteroclinic cycle, not
just a heteroclinic connection.

\vspace*{-0.6cm}
\section{Conclusion}
The unstable but recurrent coherent structures observed in turbulent
boundary layers and in transitional flows are an aspect of turbulent
flows. Invariant sets capture some features of these coherent
structures and their dynamics.  While the notion of coherent
structures varies with the means used to identify them, the notion of
invariant sets is much more precise.  Compact but linearly unstable
invariant sets in \statesp\ (such as \eqva, \reqva, \po s, partially
hyperbolic tori) are exact solutions of the Navier-Stokes equation
which correspond to sustained motions of the fluid.

  As a turbulent flow evolves, every so often we catch a glimpse of a
familiar pattern.  In some instances, turbulent dynamics visualized
in {\statesp} appears pieced together from close visitations of
equilibria connected by transient interludes. These turbulent
interludes themselves
reflect close passes to other invariant sets in
{\statesp}, such as unstable periodic orbits.
Such an approach to turbulence based on a repertoire
of recurrent spatio-temporal patterns, which would be periodic or
relative periodic orbits in {\statesp}, was proposed by
\cite{Christiansen:97} as an implementation of \cite{Hopf}'s
view that turbulent flows are ergodic trajectories in {\statesp}.  A
similar approach has been suggested by \cite{Narasimha89}, who refers
to these patterns as molecules of turbulence.

The heteroclinic orbits that we present here could be the initial
steps in
charting
an atlas of the dynamics of plane Couette flow; close
passages to equilibria could be identified with nodes of Markov graph
to give a coarse form of symbolic dynamics, and then these
heteroclinic cycles would be directed links connecting nodes of the
Markov graph.  The lower branch equilibrium $\tLB$,
along with the equilibria which connect
back to it, appear to form a part of the {\statesp} boundary dividing
two regions: one laminar the other turbulent. Turbulent trajectories
appear to be trapped between that boundary and the unstable manifolds
of the upper branch equilibrium $\tUB$, as illustrated by
\cite{GHC}.

The emergence and disappearance of these heteroclinic
connections
can also
be diagnostic. The disappearance of the $\tLB$ to $\tUB$ connection is
reminiscent of other global bifurcations occurring in simpler
dynamical systems.  For instance, in the Lorenz system a series of
such bifurcations occur as the ``Rayleigh'' number is increased
\citep{jackson89}.  For plane Couette flow,
such bifurcations could be useful for marking
the onset of turbulence.

Future work in this direction should serve to clarify such points.  It
is still not entirely clear what happens at the global bifurcations
involved in the creation and annihilation of these heteroclinic
connections.  Furthermore, lists of equilibria and of the heteroclinic
connections between them found so far should by no means
be considered exhaustive.  Further investigation of plane Couette flow
as well as other geometries will most likely turn up
other dynamically important invariant sets,
and more heteroclinic connections between them.

\noindent{\bf Acknowledgments.}
The authors thank Y. Duguet and L. van Veen for helpful discussions.
D.V. was partly supported by NSF grants DMS-0407110 and DMS-0715510.
He thanks the mathematics department of the Indian Institute of
Science, Bangalore, for its hospitality and support.
P.C., J.F.G.\ and J.H.\ thank G.~Robinson,~Jr.\ for support.
J.H.\ thanks R.~Mainieri and T.~Brown,
Institute for Physical Sciences, for partial support.
Special thanks to the Georgia Tech Student Union
which generously funded our access
to the Georgia Tech Public Access Cluster Environment (GT-PACE).

\bibliographystyle{jfm}
\bibliography{references}

\end{document}